\documentclass[a4paper,11pt]{article}

\usepackage{jheppub} 

\usepackage[T1]{fontenc} 
\usepackage{bm,latexsym,amsmath,amssymb,amsfonts,mathtools,caption}
\usepackage{amssymb, amsmath, comment}
\usepackage{color}
\usepackage{hyperref}
\usepackage{booktabs}
\usepackage{mathrsfs}
\usepackage{ulem}
\usepackage{wrapfig}
\makeatletter
\newcommand{\figcaption}[1]{\def\@captype{figure}\caption{#1}}
\newcommand{\tblcaption}[1]{\def\@captype{table}\caption{#1}}

\newcommand{\kdsf}{Kerr-AdS${}_5$ }
\makeatother
\preprint{RIKEN-iTHEMS-Report-22}
\begin{document}
\title{dS${}_4$ universe emergent from Kerr-AdS${}_5$ spacetime:\\
bubble nucleation catalyzed by a black hole}

\author[a]{Issei Koga}
\author[b]{Naritaka Oshita}
\author[a]{Kazushige Ueda}
\affiliation[a]{Department of Physics, Kyushu University, 744 Motooka, Nishi-Ku, Fukuoka 819-0395, Japan}
\affiliation[b]{Interdisciplinary Theoretical and Mathematical Sciences Program (iTHEMS), RIKEN, Wako 351-0198, Japan}

\abstract{
The emergence of a four-dimensional de Sitter (dS${}_4$) universe on an expanding bubble in the five-dimensional anti-de Sitter (AdS${}_5$) background has been suggested as a possible cosmological scenario. It is motivated by the difficulties in the realization of a stable de Sitter vacua in string theory. The bubble can be nucleated in a meta-stable pure AdS${}_5$ spacetime, but it is known that a pure AdS spacetime is non-perturbatively unstable. It means that the pure AdS${}_5$ background is an idealized situation, and in realistic situations, non-linear perturbations in AdS may lead to the formation of black holes due to the gravitational turbulent instability. To investigate how the proposed scenario works in a more realistic situation, we here study the nucleation process of a vacuum bubble in the Kerr-AdS${}_5$ spacetime. Especially we investigate conditions sufficient to ensure the nucleation of a vacuum bubble with a rotating black hole and how the black hole affects the transition rate. We find that even in the \kdsf spacetime, a quasi-dS${}_4$ expansion can be realized on the nucleated vacuum bubble without contradicting the de Sitter swampland conjectures.
}
 
\maketitle

\section{Introduction}
\label{Introduction}
The difficulty of constructing de Sitter vacua in string theory (See ref. \cite{Danielsson:2018ztv} for a review) leads to a discrepancy with cosmology. Also, the swampland conjecture \cite{Ooguri:2018wrx,Garg:2018reu,Obied:2018sgi} has revived the debate on how a de Sitter universe, modeling inflation or the Universe dominated by dark energy, can be consistent with string theory.  Recently, it was pointed out that an expanding bubble in a five-dimensional anti-de Sitter (AdS) spacetime mimics a four-dimensional de Sitter spacetime on the bubble \cite{Banerjee:2018qey,Banerjee:2019fzz,Banerjee:2020wix}. The idea of the realization of the Universe on a brane was pioneered in Ref. \cite{Randall:1999vf}. The nucleation of an expanding bubble can be realized in a meta-stable AdS${}_5$ vacuum via the Coleman de Luccia (CdL) transition \cite{Coleman:1980aw}. The nucleated bubble mediates two different AdS${}_5$ vacua, and the interior vacuum has a smaller vacuum energy density.

On the other hand, a pure AdS vacuum is non-linearly unstable and may eventually form a black hole due to turbulent instability \cite{Bizon:2011gg,Dias:2011ss}. AdS spacetime has a boundary and acts like a confining box. Therefore, any finite excitation in the AdS would not be dissipated and might be expected to explore all possible configurations, which may eventually lead to the formation of small black holes. For example, a rotating black hole is expected to form as a result of gravitational turbulent instability in AdS \cite{Dias:2011ss}. To see if the proposed scenario \cite{Banerjee:2018qey,Banerjee:2019fzz,Banerjee:2020wix} is applicable even to a \kdsf black hole, it is important to study if we can construct the solution of an expanding bubble nucleated via the vacuum decay in \kdsf spacetime. The nucleation of a vacuum bubble surrounding a black hole has been actively studied (see, e.g., Refs. \cite{Gregory:2013hja,Burda:2015yfa,Oshita:2019jan,Oshita:2020ksb,Gregory:2020cvy,Hayashi:2020ocn,Gregory:2020hia,Shkerin:2021zbf,Saito:2021vut,Saito:2022hty}), mainly in the motivation of the Higgs metastability \cite{Degrassi:2012ry,Buttazzo:2013uya}.

\begin{figure}[t]
\centering
 \includegraphics[width=15cm]{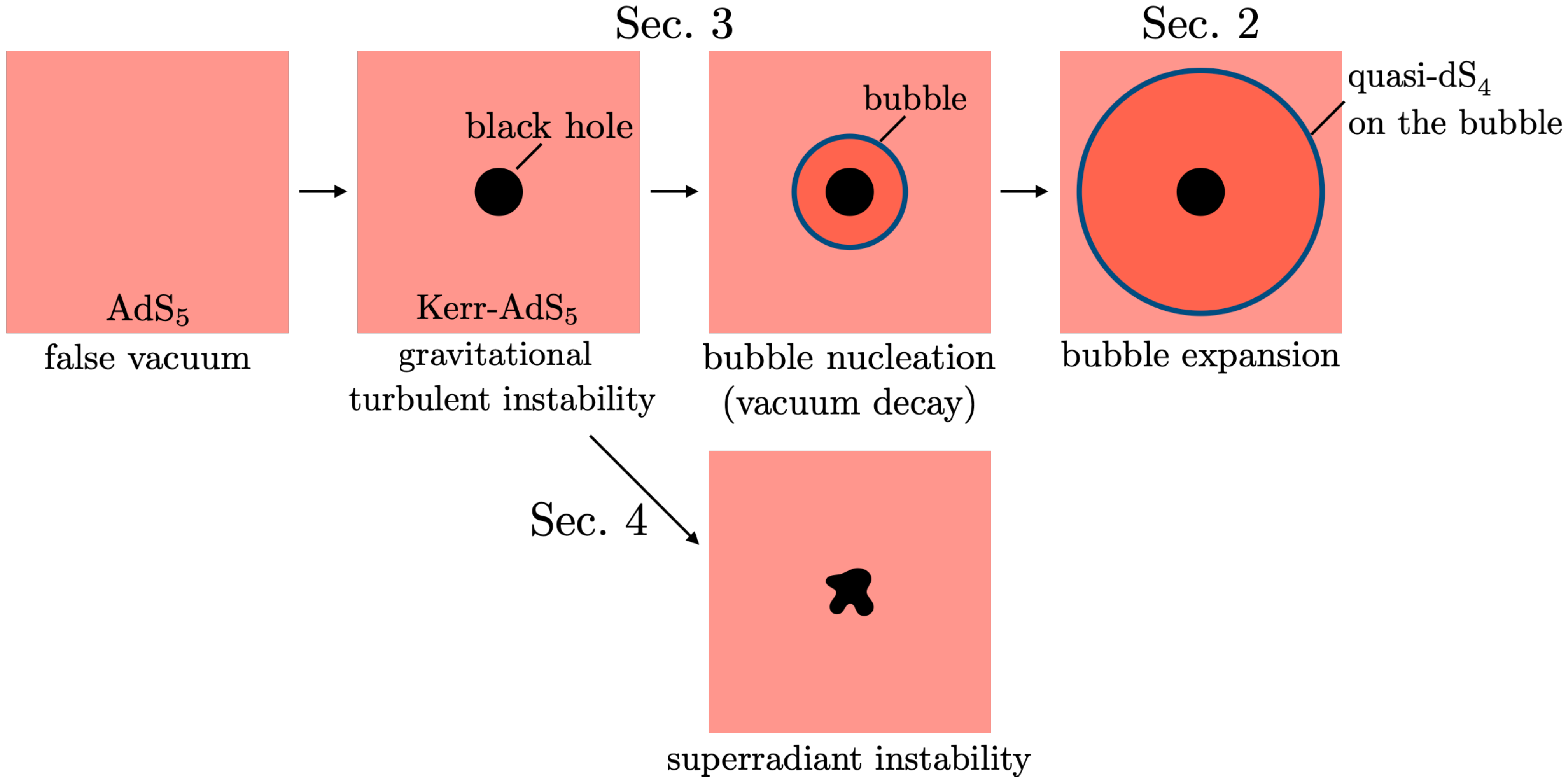}
  \caption{Schematic picture of the cosmological scenario proposed here.}
\label{schematic}
\end{figure}
In this paper, we study the nucleation process of a vacuum bubble in the \kdsf spacetime with the restriction of equal rotations. We here assume a thin-wall bubble that separates two different vacua and whose dynamics is governed by the Israel junction conditions \cite{Israel:1966rt}. We then compute the on-shell Euclidean action of a bubble solution, $S_{\rm E}$, which gives the exponent of a transition amplitude. We find that the semi-classical approximation is broken with $S_{\rm E}< 0$ for a massive and rapidly spinning seed black hole. We also derive the Friedmann-like equation describing the expansion of the four-dimensional universe realized on the bubble or brane. It is well known that the mass of the black hole acts like {\it radiation} on the brane, called dark radiation \cite{Ida:1999ui,Mukohyama:1999qx}. We also find that the rotations of the \kdsf black hole give a term of $\sim 1/(\text{scale factor})^6$ in the Friedmann-like equation. In addition, the anisotropy originating from the rotations in the induced metric on the bubble decays in $1/(\text{scale factor})^4$. On the other hand, the rotations of the seed black hole lead to the superradiant instability. Thus, our computation and results are applicable only when the time scale of the supperradiant instability is much longer than that  of the vacuum metastability, and it depends on the parameters of the false vacuum state.
We then compare the decay rate of the false vacuum with the instability rate of the superradiance in the \kdsf background and reveal the parameter region of the seed \kdsf black hole where the condition is satisfied.

This paper is organized as follows. In Sec. \ref{formalism}, we briefly review the Israel junction conditions governing the dynamics of the bubble by following Ref. \cite{Rocha:2015tda} in which the Israel junction conditions were applied to a thin shell in the \kdsf spacetime. Then we derive the Friedmann-like equation from the Israel junction conditions. We then show the formalism to compute the on-shell Euclidean action of the thin wall. In Sec. \ref{ds4_on_bubble}, we show the condition of the bubble nucleation in the \kdsf background. We then estimate the decay rate by employing the Euclidean path integral to see the most probable transition process. As a result, we find that the stationary solution, in which the Euclidean bubble does not oscillate in the radial direction, gives the most probable process of vacuum decay. Also, we find the breakdown of the semi-classical approximation for a rapidly spinning and massive black hole. In Sec. \ref{QNM_part}, we compare the lifetime of the false vacuum, caused by the quantum mechanical instability, with that of the seed \kdsf black hole unstable due to the (classical) superradiance. Sec. \ref{conclusions} is devoted to the conclusions. Figure \ref{schematic} illustrates the outline of our paper along with some schematic pictures. Throughout the paper, we take the natural units and the five-dimensional Newton's constant is set to $G_5 = 1$.

\section{Formalism} 
\label{formalism}
\subsection{Dynamics of a bubble in the \kdsf spacetime}
Based on the Euclidean path integral, one can estimate the transition amplitude from a false vacuum state to a true vacuum one by computing the difference of the on-shell Euclidean action before and after the transition. In the thin-wall approximation, the dynamics of a meta-stable field is described by that of a thin-wall bubble. We are interested in the motion of a thin wall that mediates two different \kdsf spacetime, which was studied in Ref. \cite{Delsate:2014iia}. Here we review their analysis that employs the Israel junction conditions to describe the thin-shell dynamics in \kdsf spacetime.

\begin{figure}[htbt]
\centering
 \includegraphics[width=11cm]{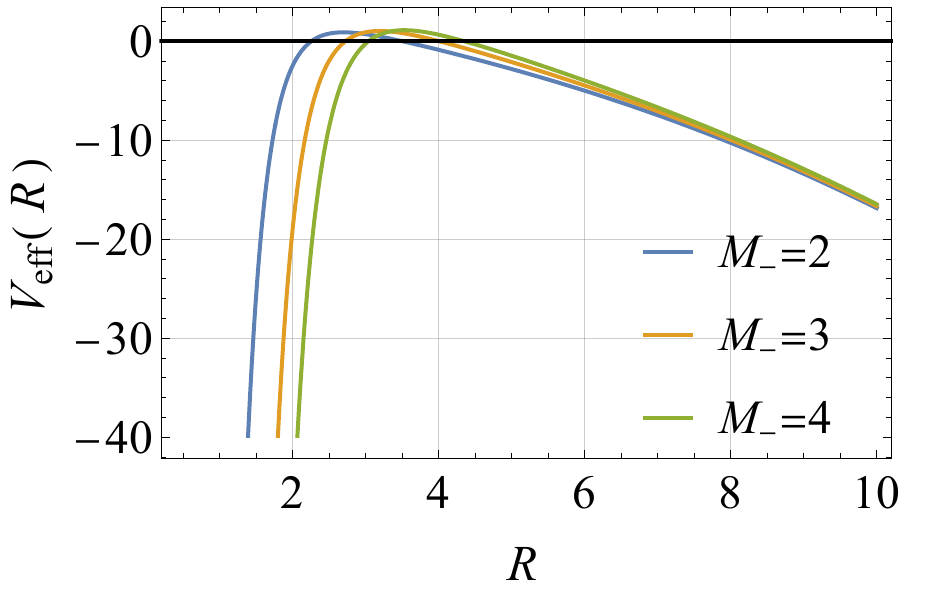}
  \caption{Plot of $V_{\rm eff}$ for $M_+=1$, $a_+=0.6$, $l_+=7$, $l_-=4$, $m_0=500$, and $w=-1$. The remnant mass is set to $M_-=2$, $3$, and $4$.}
\label{potential_01}
\end{figure}

The \kdsf spacetime has two rotations, and we consider the case of equal rotations for simplicity. Given the cosmological constant, $\Lambda (\equiv -6 l^2)$, spin, $a$, and mass parameter, $M$, of a black hole, the line element on the \kdsf spacetime is expressed by the coordinates of $x^\mu = (t,r,\theta,\psi,\phi)$ \cite{Gibbons:2004js}:
\begin{eqnarray}
ds^2=-f(r)^2 dt^2+g(r)^2dr^2+r^2{\hat g}_{ab} dx^a dx^b
+h(r)^2 \bigl[ d\psi+A_a dx^a -\Omega(r) dt   \bigr]^2,
\label{kdsf_metric}
\end{eqnarray}
where
\begin{eqnarray}
&&A_a dx^a \equiv \frac{1}{2} \cos\theta d\phi,~g(r)^2 \equiv \Biggl(
1+\frac{r^2}{l^2}-\frac{2M\Xi}{r^2}+\frac{2 M a^2}{r^4}
\Biggr)^{-1},~h(r)^2 \equiv r^2\Biggl(
1+\frac{2Ma^2}{r^4}
\Biggr),~\nonumber\\
&&\Omega(r) \equiv \frac{2Ma}{r^2 h(r)^2},~
f(r)\equiv\frac{r}{g(r) h(r)},~\Xi \equiv 1-\frac{a^2}{l^2},~\hat{g}_{ab}dx^a dx^b \equiv \frac{1}{4}(d\theta^2+\sin^2\theta d\phi^2).\nonumber
\end{eqnarray}
We then set a junction surface $\Sigma=\{
x^{\mu}:t=T(\tau),~r=R(\tau)
\}$ that separates the interior and exterior spacetime in which the cosmological constant is $\Lambda_-=-6 l_-^{-2}$ and $\Lambda_+=-6 l_+^{-2}$, respectively. In the following, the superscript or subscript of $+$ ($-$) denotes the exterior (interior) quantities. Given the junction surface, one can define the induced metric on the inner and outer surface by $q^{(-)}_{ab}$ and $q^{(+)}_{ab}$, respectively. Also, one can define the extrinsic curvature $K^{(\pm)}_{ab}$ on each surface. The first and second Israel junction conditions are given by
\begin{eqnarray}
&&q_{ij}^{(+)}=q_{ij}^{(-)}=q_{ij}, \label{1stJunction}\\
&&(K_{ij}^{(+)}-K_{ij}^{(-)})-q_{ij}(K^{(+)}-K^{(-)})
=-8\pi S_{ij},
\label{2ndJunction}
\end{eqnarray}
where $K^{(\pm)} = q^{ab} K_{ab}^{(\pm)}$ and $S_{ij}$ is the reduced energy momentum tensor of the thin wall. Going to the comoving frame with the following coordinate transformations:
\begin{align}
d\psi &\to d\psi + \Omega_{\pm} ({R}(t)) dt,
\label{psi_redefine}\\
dt_{\pm} &\to \frac{d {T}_{\pm}}{d \tau} d\tau,\\
dr &\to \frac{d {R}}{d \tau} d\tau,
\end{align}
the original metric in (\ref{kdsf_metric}) reduces to the induced metric, $q^{(\pm)}_{ij}$, as
\begin{equation}
ds_{\pm}^2 =q^{(\pm)}_{ij} dy^i dy^j= - \left[ f_{\pm}^2 \left( \frac{d {T}_{\pm}}{d \tau} \right)^2 - g_{\pm}^2 \left( \frac{d {R}}{d \tau} \right)^2 \right] d\tau^2 + r^2 \hat{g}_{ab} dx^a dx^b + h_{\pm}^2 \left[ d \psi + A_a dx^a \right]^2,
\label{induced_metric_qij}
\end{equation}
where $\tau$ is the proper time on the thin wall.
From the condition of the comoving frame and the first Israel junction condition (\ref{1stJunction}), we have
\begin{align}
&f_{\pm}^2 \dot{ T}_{\pm}^2 - g_{\pm}^2 \dot{ R}^2 = 1,\\
&M_+ a_+^2 = M_- a_-^2 \equiv Ma^2,
\end{align}
where a dot represents the derivative with respect to the proper time $\tau$.
Note that the junction surface we take is not spherical although it locates at $r = {R}(\tau)$ in our coordinates. In the proper length, the surface is axisymmetric and is deformed from a spherical shape. Therefore, it is natural to regard the thin wall as an imperfect fluid with anisotropic components \cite{Delsate:2014iia}:
\begin{equation}
{\cal S}_{ij} = (\sigma + P) u_i u_j + P q_{ij} +2 \varphi u_{(i} \xi_{j)} + \Delta P {R}^2 \hat{g}_{ij},
\label{emt_wall}
\end{equation}
where
\begin{eqnarray}
\xi = h^{-1} ({R}) \partial_{\psi}, \ u = \partial_{\tau}.
\end{eqnarray}
The extrinsic curvature, $K_{ij}^{(\pm)}$, is
\begin{equation}
K^{(\pm)}_{ij} = e^{\mu}_{i} e^{\nu}_j K^{(\pm)}_{\mu\nu} \ \text{with} \ K^{(\pm)}_{\mu\nu}=(g_{\mu\sigma}-n_\mu n_\sigma)\nabla^\sigma n_\nu,
\end{equation}
where $e^{\mu}_i \equiv dx^{\mu}/dy^i$ and $n_{\mu}$ is the unit normal vector on the wall
\begin{equation}
n_\pm= f_\pm g_\pm
(-\dot{R}dt_\pm+\dot{T}_\pm dr_\pm ).
\end{equation}
Then the second Israel junction condition reads \cite{Delsate:2014iia}
\begin{align}
\sigma &= -\frac{(\beta_+ - \beta_-) ({R}^2 h)'}{8 \pi {R}^3},\\
P &= \frac{h}{8 \pi {R}^3} [{R}^2 (\beta_+ - \beta_-)]',\\
\varphi &= \frac{(\Omega_+' - \Omega_-') h^2}{16 \pi {R}},\\
\Delta P &= \frac{(\beta_+ - \beta_-)}{8 \pi} \left[ \frac{h}{R} \right]',
\end{align}
where a prime denotes $d/d{R}$, $\beta_{\pm} \equiv f_{\pm}^2 \dot{T}_{\pm} = \epsilon_{\pm} f_{\pm} ({R}) \sqrt{1+g_{\pm}^2 \dot{R}^2}$, and $\epsilon_{\pm}$ is the sign of $\dot{T}_{\pm}$. Combining those conditions along with an extra condition for the fluid, i.e., the equation of state of the thin wall
\begin{equation}
P = w \sigma,
\end{equation}
we have the following equation
\begin{equation}
\frac{[{R}^2 (\beta_+ - \beta_-)]'}{{R}^2 (\beta_+ - \beta_-)} = -w \frac{[{R}^2 h]'}{{R}^2 h}.
\end{equation}
It reduces to the following equation that governs the dynamics of the thin wall
\begin{equation}
\beta_+ - \beta_- = - \frac{m_0^{1+3w/2}}{{R}^{2 (1+w)} h^w ({R})} \equiv -b(R),
\end{equation}
where $m_0$ is an integration constant and has the mass dimension\footnote{Note that we here take $G_5=1$ that has the dimension of $1/(\text{mass})^3 = (\text{length})^3$. Explicitly showing the constant $G_5$, we have $m_0 \to G_5 m_0 = (\text{length})^2$ which makes the function $b(r)$ non-dimensional.}. We finally obtain the (integrated) equation of motion of the thin wall
\begin{eqnarray}
\dot{R}^2+V_{\rm eff}=0,
\label{Lorentzian_motion}
\end{eqnarray}
where $V_{\rm eff}$ is the effective potential of the thin wall
\begin{eqnarray}
V_{\rm eff}=\frac{1}{g_-^2}
\Biggl[
1-\biggl(
\frac{-f_+^2+f_-^2+b^2}{2bf_-}
\biggr)^2
\Biggr].
\end{eqnarray}
The effective potential depends on the mass parameter ($M_{\pm}$), spin parameter ($a_{\pm}$), AdS radius ($l_{\pm}$), and $m_0$. As an example, Figure \ref{potential_01} shows some effective potentials, and one can see that for each potential, there exists a forbidden region in which $\dot{R}^2 < 0$.
Assuming the brane is dominated by the effective potential of the meta-stable field, we set the equation-of-state parameter as $w=-1$ and we have
\begin{eqnarray}
&&\biggl(\frac{\dot{R}}{R}\biggr)^2+R^{-2}=
-\frac{1}{4}
\left(
-\frac{(l_+^2-m_0)^2}{l_+^4 m_0} 
-\frac{m_0}{l_-^4}
+\frac{2}{l_-^2}
+\frac{2m_0}{l_-^2 l_+^2}
\right)
\nonumber\\
&&~~~~~~~~~~
-M_-a_-^2 \bigl(
-\frac{1}{a_-^2}-\frac{1}{a_+^2}+\frac{1}{l_-^2}+\frac{1}{l_+^2}
-\frac{1}{m_0}
\bigr)R^{-4}
-2Ma^2 R^{-6}
+\frac{(Ma^2)^2}{m_0} R^{-8}
\nonumber\\
&&~~~~~~~~~~~~~
-\frac{ (a_--a_+)(a_-+a_+)(l_--l_+)(l_-+l_+)M_- m_0 }{ a_+^2l_-^2l_+^2(2Ma^2+R^4) }
+\frac{(a_-^2-a_+^2)^2 M_-^2 m_0}{a_+^2(2Ma^2+R^4)^2}
\nonumber\\
\label{FriedmanLike}
\end{eqnarray}
The equation of (\ref{FriedmanLike}) can be compared with the Friedmann equation of a closed universe
\begin{align}
& \frac{(d R_s/dt)^2}{R_s^2}
+\frac{1}{R_s^2}=\frac{8\pi G}{3}\rho +\frac{\Lambda}{3},\\
\text{with} \ &ds^2 = -dt^2 + {R}_s^2(t) \left[dr^2 + \sin^2 r (d\theta^2 + \sin^2 \theta d\phi^2) \right],
\end{align}
where ${R}_s$ is the scale factor and $G$ is the four-dimensional Newton's constant. Identifying the radius of the bubble wall ${R}$ with the scale factor ${R}_s$, the equation (\ref{FriedmanLike}) can be regarded as the Friedmann equation. Indeed, the induced metric, $q_{ij}$, has the form
\begin{align}
\begin{split}
ds^2 &= q_{ij} dy^i dy^j = -d\tau^2 + {R}^2 (\tau) \left[ \hat{g}_{ab} dx^a dx^b + \left( 1+ \frac{2Ma^2}{{R}^4} \right) (d \psi + A_a dx^a)^2 \right],\\
&= -d\tau^2 + {R}^2 (\tau) \left[ d\chi^2 + \sin^2 \chi (d\zeta^2 + \sin^2 \zeta d\xi^2) \right] +  \frac{2Ma^2}{{R}^2} (d \psi + A_a dx^a)^2,
\end{split}
\label{FLRW_metic}
\end{align}
where we used the fact that the metric on $S^3$ sphere is described by the combination of the Fubini-Study metric and the K\"{a}hler metric, $A_a dx^a$, along with $\psi$ \cite{Gibbons:2004uw}
\begin{equation}
\hat{g}_{ab} dx^a dx^b + (d \psi + A_a dx^a)^2 = d\chi^2 + \sin^2 \chi (d\zeta^2 + \sin^2 \zeta d\xi^2).
\end{equation}
Note that the last term in (\ref{FLRW_metic}) is of the order of ${\cal O}({R}^{-2})$ and it becomes negligible as the bubble, i.e., an emergent four-dimensional universe, expands. Therefore, in the case of $Ma^2/{R}^4 \ll 1$, the induced metric reduces to the Friedmann–Lema\^{i}tre–Robertson–Walker (FLRW) metric. One can read some intriguing features from the Friedmann-like equation in (\ref{FriedmanLike}):
\begin{itemize}
    \item The asymptotic behaviour of the expansion is equivalent to the (four-dimensional) de Sitter expansion.
    \item In the Friedmann-like equation and the FLRW-like metric, the effect of the rotations of the black hole is diluted in $1/{R}_s^6$ and in $1/{ R}_s^4$, respectively, in the limit of ${R}_s \to \infty$.
    \item The curvature of the space is positive.
\end{itemize}

\subsection{Euclidean action}
The nucleation rate of a vacuum bubble $\Gamma$ is given by
\begin{equation}
\Gamma \simeq A e^{-B}.
\end{equation}
The prefactor $A$ has the dimension of $(\text{time})^{-1}$ and is determined by the zero modes and loop corrections of the Euclidean solution. The exponent $B$ determines the transition amplitude and is given by the difference of the on-shell Euclidean action before and after the phase transition:
\begin{align}
&B=S_{\rm E} (\phi_{\rm bubble}) - S_{\rm E} (\phi_{\rm false}),
\end{align}
where $\phi_{\rm bubble}$ is the configuration of a nucleated bubble and $\phi_{\rm false}$ is the trivial solution of a false vacuum state. We compute the Euclidean action by following the procedures in Ref. \cite{Gregory:2013hja} (see Ref. \cite{Koga:2019yzj} for the case of five-dimensional background). The transition rate is mainly determined by the factor $B$, and therefore, we will concentrate on the estimation of $B$ in this paper, and the prefactor $A$ will be determined by the dimensional analysis later. The Euclidean space of the \kdsf spacetime is obtained by performing the Wick rotation, $t \to -i t_{\rm E}$, which leads to a complex metric\footnote{For the convenience of computation, one can perform the analytic continuation of $a \to i a$, which makes the metric real. Note that in this scheme, one should make sure that the analytically continued spin $a$ is continued back to the real spin after getting the transition amplitude for the imaginary spin.} \cite{Wu:2004db}. The Wick rotation of the proper time, $\tau \to -i\tau_{\rm E}$, also changes the integrated equation of motion of the thin wall (\ref{Lorentzian_motion}) as
\begin{equation}
\dot{R}^2 - V_{\rm eff} = 0.
\label{Euclidean_motion}
\end{equation}

The whole configuration after the phase transition can be divided into four parts: the region near the black hole horizon, ${\cal H}$, bubble wall, ${\cal W}$, and the interior and exterior regions, ${\cal M}_-$ and ${\cal M}_+$, respectively. The on-shell action of the final state, $S_{\rm E}^{\rm (f)}$, is then decomposed as
\begin{equation}
S_{\text{E}}^{\rm (f)} = S_{\cal H_-} + S_{\cal M_-} +S_{\cal M_+} + S_{\cal W},
\end{equation}
where
\begin{align}
S_{\cal H_\pm} &\equiv - \frac{1}{16 \pi} \int_{\mathcal H}dx^5 \sqrt{g_{\rm E}}~
{\mathcal R} + \frac{1}{8\pi} \int_{\partial \mathcal H} dx^4 \sqrt{q_{\rm E}} \tilde{K}_{\text{E}},\\
S_{{\cal M}_\pm} &\equiv - \frac{1}{16 \pi} \int_{\mathcal M_{\pm}} dx^5 \sqrt{g_{\rm E}}~ {\mathcal R} - \int_{\mathcal M_{\pm}}
dx^5 \sqrt{g_{\rm E}}~{\mathcal L}_m^{\rm (on-shell)} 
+\frac{1}{8\pi} \int_{\partial \mathcal M_{\pm}} dx^4 \sqrt{q_{\rm E}}
\tilde{K}_{\text{E} \pm},\\
S_{\cal W} &\equiv - \int_{\mathcal W} dx^5 \sqrt{g_{\rm E}}~{\mathcal L}_m^{\rm (on-shell)} = \int_{\mathcal W} dx^4 \sqrt{q_{\rm E}} (\sigma-\Delta P), \label{wall_action}
\end{align}
where ${\cal L}_m^{\rm (on-shell)}$ is the on-shell action of the fluid on the wall. The off-shell action and the detailed derivation of ${\cal L}_m^{\rm (on-shell)} = -(\sigma -\Delta P)$ is presented in Appendix \ref{app_onshell}.
We here define the unit normal vector on the wall, $\tilde{n}_{\pm \mu}$, and unit normal vector on a $t_{\rm E}$-constant surface, $\tilde{u}_{\pm}^{\mu}$, as
\begin{align}
    &\tilde{n}_{\pm \mu}=\pm f_{\pm}(r) g_{\pm}(r) \biggl(
    -\frac{d {R}}{d \tau_{\rm E}},\frac{d {T}_{{\rm E}\pm}}{ d\tau_{\rm E}},0,0,0
    \biggr),\\
    &\tilde{u}^{\mu}_{\pm} = \biggl(
    \frac{1}{f_{\pm}},0,0,-\frac{i\Omega_{\pm}}{f_{\pm}},0
    \biggr),
\end{align}
and $\tilde{K}_{\rm E \pm}$ is the trace of the extrinsic curvature associated with $\tilde{n}_{\pm \mu}$.
The contribution of the black hole horizon to the Euclidean action is \cite{Wu:2004db}
\begin{align}
S_{\cal H_\pm} =- \frac{\pi^2(r_\pm^2+a_{\pm}^2)^2}{2r_\pm(1-a_{\pm}^2/l_\pm^2)^2}=-\frac{\cal A_\pm}{4},
\end{align}
where ${\cal A_\pm}$ is the horizon area of the remnant black hole \cite{Gibbons:2004js,Gibbons:2004uw}, and this is nothing but the Bekenstein-Hawking entropy \cite{Hawking:1998kw} multiplied by $-1$. On the other hand, from the Hamiltonian constraints and having the Killing coordinate $t_{\rm E}$, the bulk component is
\begin{align}
S_{{\cal M}_\pm} & =  \frac{1}{8 \pi} \int_{\mathcal W}dx^4 \sqrt{q_{\rm E}}~
\kappa_\pm
+\frac{1}{8 \pi} \int_{\mathcal W} dx^4 \sqrt{q_{\rm E}}~ \tilde{K}_{\text{E} \pm},
\end{align}
where $\kappa_\pm$ is the non-zero surface term that originates from the Ricci scalar $\kappa_\pm \equiv \tilde{n}_{\pm \nu} \tilde{u}_{\pm}^{\mu} \nabla_{\mu} \tilde{u}_{\pm}^{\nu}$.
This surface term reads
\begin{align}
    \kappa_\pm = \tilde{n}_{\pm \nu} \tilde{u}_{\pm}^{\mu} \nabla_{\mu} \tilde{u}_{\pm}^{\nu}
    =\pm \dot{T}_{{\rm E} \pm} \frac{R}{h^2 g_{\pm}^3} \left[h g_{\pm}' + \frac{g_{\pm}}{R} \left(R h' - h \right) \right] = \mp \dot{T}_{{\rm E}\pm}\frac{f'_{\pm}}{g_{\pm}}.
\end{align}
Thus, the Euclidean action, $S_{\rm E}^{\rm (f)} \equiv S_{\rm E} (\phi_{\rm bubble})$, is 
\begin{align}
\begin{split}
&S_{\rm E}^{\rm (f)}=-\frac{\cal A_-}{4} +\frac{1}{8 \pi} \int_{\cal W} dx^4 \sqrt{q_{\rm E}} ({K}_{\text{E} +} - {K}_{\text{E} -})+ \int_{\cal W} dx^4 \sqrt{q_{\rm E}} (\sigma-\Delta P)\\
&+\frac{1}{8\pi} \int_{\cal W} dx^4 \sqrt{q_{\rm E}}(\kappa_+ +\kappa_-),
\end{split}\\
&= -\frac{\cal A_-}{4} + \int_{\cal W} dx^4 \sqrt{q_{\rm E}} \left( -\frac{1}{3} \sigma - \frac{1}{3} \Delta P \right) + \frac{1}{8 \pi} \int_{\mathcal W} dx^4 \sqrt{q_{\rm E}} (\kappa_+ + \kappa_-),\\
&= -\frac{\cal A_-}{4} + \oint d\tau_{\rm E} \int d^3 x \frac{\sqrt{q_{\rm E}}}{8 \pi}  \left\{ \epsilon_+ \frac{h f_+/R-h f'_+}{R} \sqrt{1 -V_{\rm eff} g_+^2} - (+ \leftrightarrow -) \right\},\label{total_action}
\end{align}
where we used $\tilde{K}_{\pm} = \pm K_{\pm}$ and the second junction condition (\ref{2ndJunction}) in the first and second equalities.
Finally, the exponent of the transition amplitude is
\begin{equation}
B = S_{\rm E}^{\rm (f)} - S_{\rm E}^{\rm (i)} = - \Delta S_{\rm BH} + \oint d\tau_{\rm E} \int d^3 x \frac{\sqrt{q_{\rm E}}}{8 \pi}  \left\{ \epsilon_+ \frac{h f_+/R-h f'_+}{R} \sqrt{1 -V_{\rm eff} g_+^2} - (+ \leftrightarrow -) \right\},
\label{on_shell_action_fullform}
\end{equation}
where $S_{\rm E}^{\rm (i)} \equiv S_{\rm E} (\phi_{\rm false}) = - {\cal A_+}/4$ and $\Delta S_{\rm BH} \equiv ({\cal A_-} - {\cal A_+})/4$.

\section{Nucleation of a dS${}_4$ universe in \kdsf background}
\label{ds4_on_bubble}
In this section, we numerically compute the on-shell Euclidean action that is the exponent of the transition amplitude and clarify which parameters of the thin-wall model describe bubble nucleation.

\subsection{Parameter region for the nucleation of a vacuum bubble}\label{secbounce1}

\begin{figure}[b]
\centering      
 \includegraphics[width=15cm]{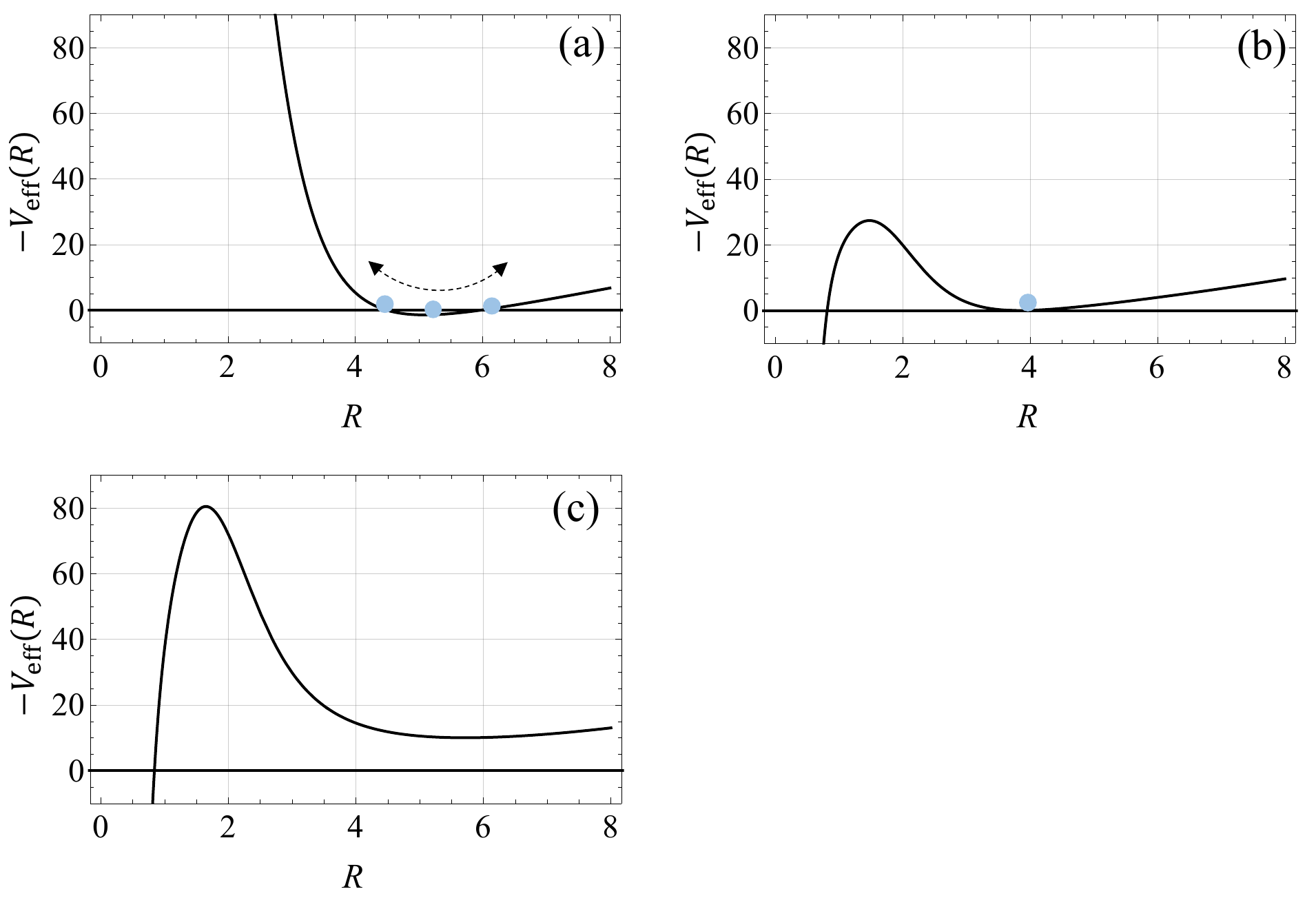}
  \caption{Plots of the effective Euclidean potential $-V_{\rm eff}(R)$. (a) is a non-stationary solution, (b) is a stationary solution and a potential (c) does not have a solution of a bubble nucleation. The parameters are set to $a_+=1$ and $M_-=14.04$. The mass of a seed black hole is set to (a) $M_+= 1$, (b) $M_+= 10$, and (c) $M_+= 19.5$.}
\label{eplot}
\end{figure}

We here compute the transition amplitude, $e^{-B}$, by substituting the Euclidean solution into (\ref{total_action}). In the thin-wall model, we have several parameters: $m_0, a_\pm, M_\pm$, and $l_\pm$. Depending on the parameters, we have two types of solutions to the integrated equation of motion (\ref{Euclidean_motion}): a stationary and a non-stationary solution. For a non-stationary solution, the bubble in the imaginary time oscillates between the two largest single roots of $V_{\rm eff}$ (Figure \ref{eplot}-(a)). For a stationary solution, the nucleated bubble has no radial velocity, $\dot{R} = 0$, and stands at the largest double root of $V_{\rm eff}$ (Figure \ref{eplot}-(b)). Also, depending on the parameters in the model, there are neither stationary nor non-stationary solutions, which means one cannot construct the semi-classical solution describing the corresponding bubble nucleation (see Figure \ref{eplot}-(c)).

\begin{figure}[t]
\centering
 \includegraphics[width=8cm]{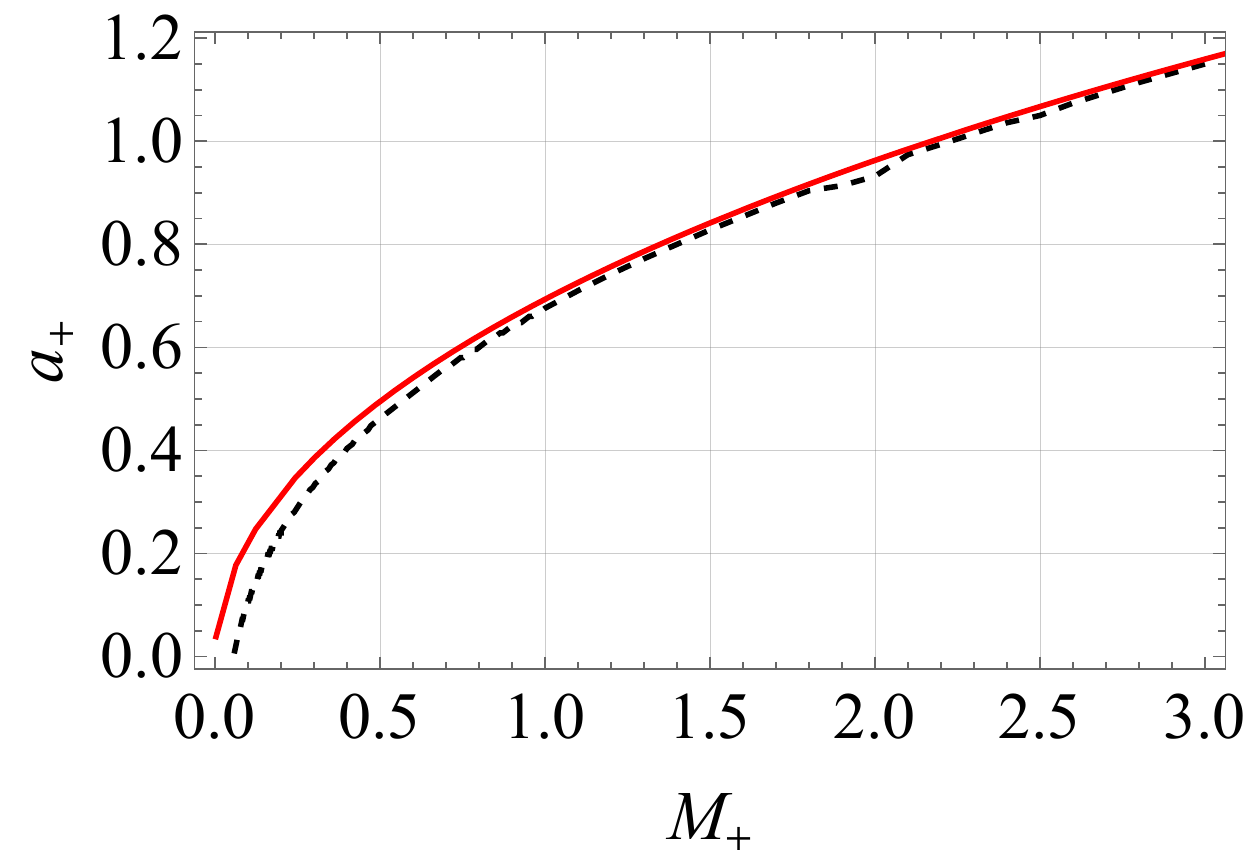}
  \caption{Constraints on the parameter space of $(M_+, a_+)$ from the condition that there is no naked singularity before (below the red solid line) and after (below the black dashed line) the nucleation of a stationary vacuum bubble.
  }
\label{bound_bh}
\end{figure}

We first numerically study the allowed parameter range for the existence of stationary solutions. We set $l_+=7, l_-=4, m_0=500$ in the following computations. Fixing $l_{\pm}$ and $m_0$ is equivalent to setting an effective potential of a metastable field that has the false and true vacuum states of the AdS radii $l_{\pm}$ and a potential barrier leading to the thin wall with the tension $m_0$. We then put constraints on the parameters of a seed black hole, $a_+$ and $M_+$, by imposing the condition that the seed and remnant black holes are regular, i.e., no naked singularity, as is shown in Figure \ref{bound_bh}. From the plot, one can read that the condition that the remnant black hole is regular is more stringent than that of the regularity of the seed black hole.
\begin{figure}[t]
\centering
\includegraphics[width=10cm]{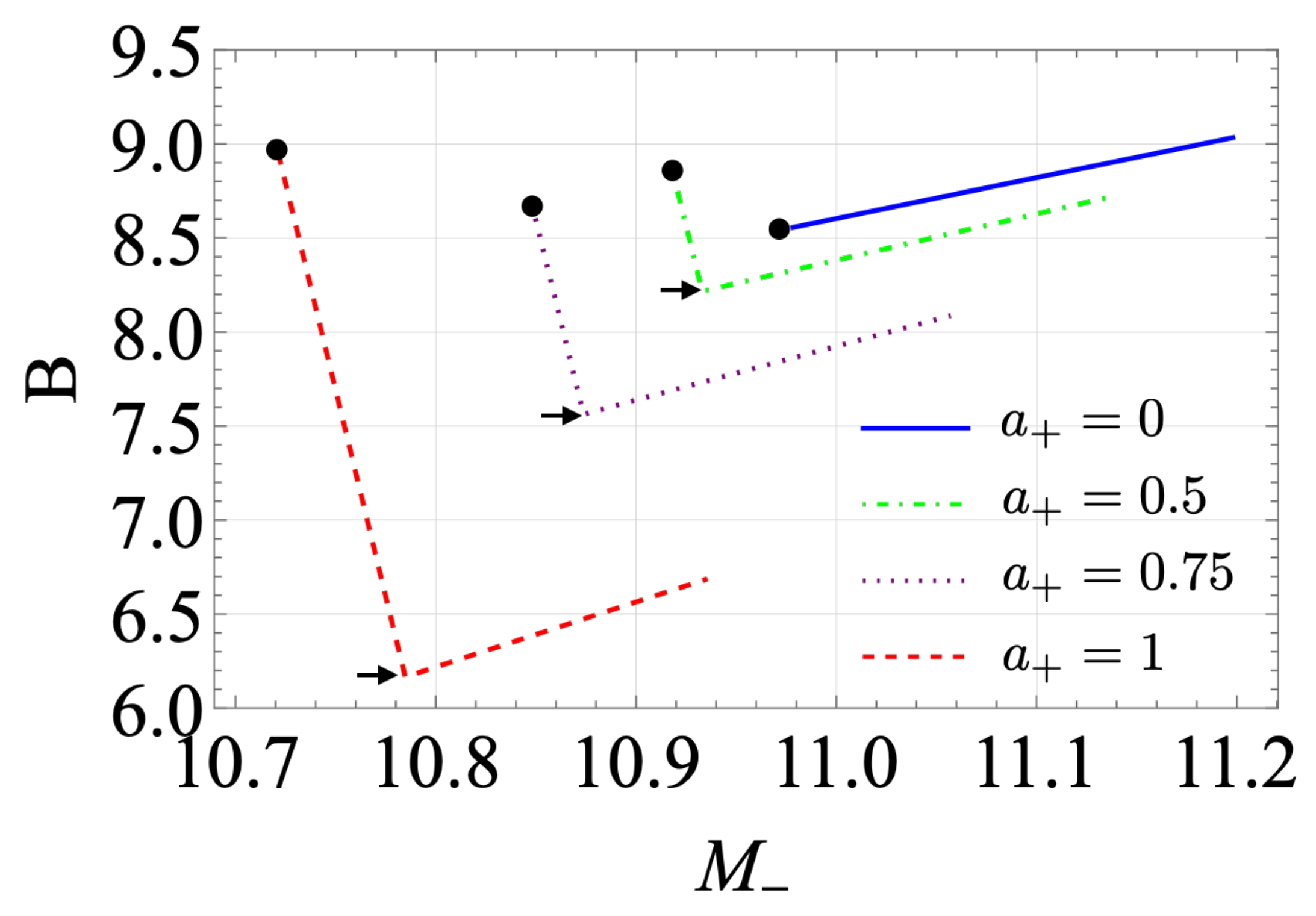}
\caption{Plots of $B$ for stationary (black markers) and non-stationary (lines) solutions with various spin parameters $a_+$ with $M_+ =8$. Arrows indicate the points where the integration part in (\ref{on_shell_action_fullform}) vanishes and $B=-\Delta S_{\rm BH}$ holds. We here show the values of $B$ for $10^{-13} \leq \delta \leq 0.02$, where $\delta \equiv M_-/M_-^{\rm crit} -1$, as $\delta \gtrsim 0.02$ in our parameter set leads to the negative sign of $\epsilon_{\pm}$ that alters the orientation of the bubble wall. The value of $M_-$ at the stationary point increases as the spin decreases, and approaches to the value of $M_- \simeq 10.978$ which is the asymptotic value for the seed AdS-Schwarzschild black hole in our parameter set.}
\label{displace_plot}
\end{figure}
In the allowed region, we can take the parameter sets in which Euclidean solutions, describing bubble nucleation, exist. We perform the computation of the Euclidean action (\ref{total_action}) to estimate the transition amplitude and find that the stationary solution exists at the minimum value of $M_{-}$ as shown in Figure \ref{displace_plot}. For a non-rotating black hole ($a_+ = 0$), the stationary solution gives the least action, i.e., the highest transition amplitude. For $a_+> 0$, on the other hand, the action at the stationary solution does not lead to the least action. The least action can be found at another solution where the integration part in (\ref{on_shell_action_fullform}) vanishes. The integration includes gravity on the bubble (i.e., curvature $K_{{\rm E}\pm}$ and surface gravity $\kappa_{\pm}$) and the anisotropic wall tension (i.e., $P+ \Delta P$) and can be zero due to the balance between gravity and the anisotropic tension. For a spherical black hole, it vanishes at the stationary point \cite{Gregory:2013hja}. We find that when the rotation is involved and the bubble wall is anisotropic imperfect fluid, the solution for which the integration part in (\ref{on_shell_action_fullform}) vanishes can be non-stationary. This non-trivial behaviour may be determined by the complicated balance among gravity, anisotropic wall tension, and rotations. We leave further discussion to future work. In such a case, the transition amplitude of the most probable nucleation process is governed only by the Bekenstein-Hawking entropy
\begin{equation}
e^{-B} = e^{\Delta S_{\rm BH}}.
\end{equation}

\begin{figure}[htbt]
\centering
 \includegraphics[width=10cm]{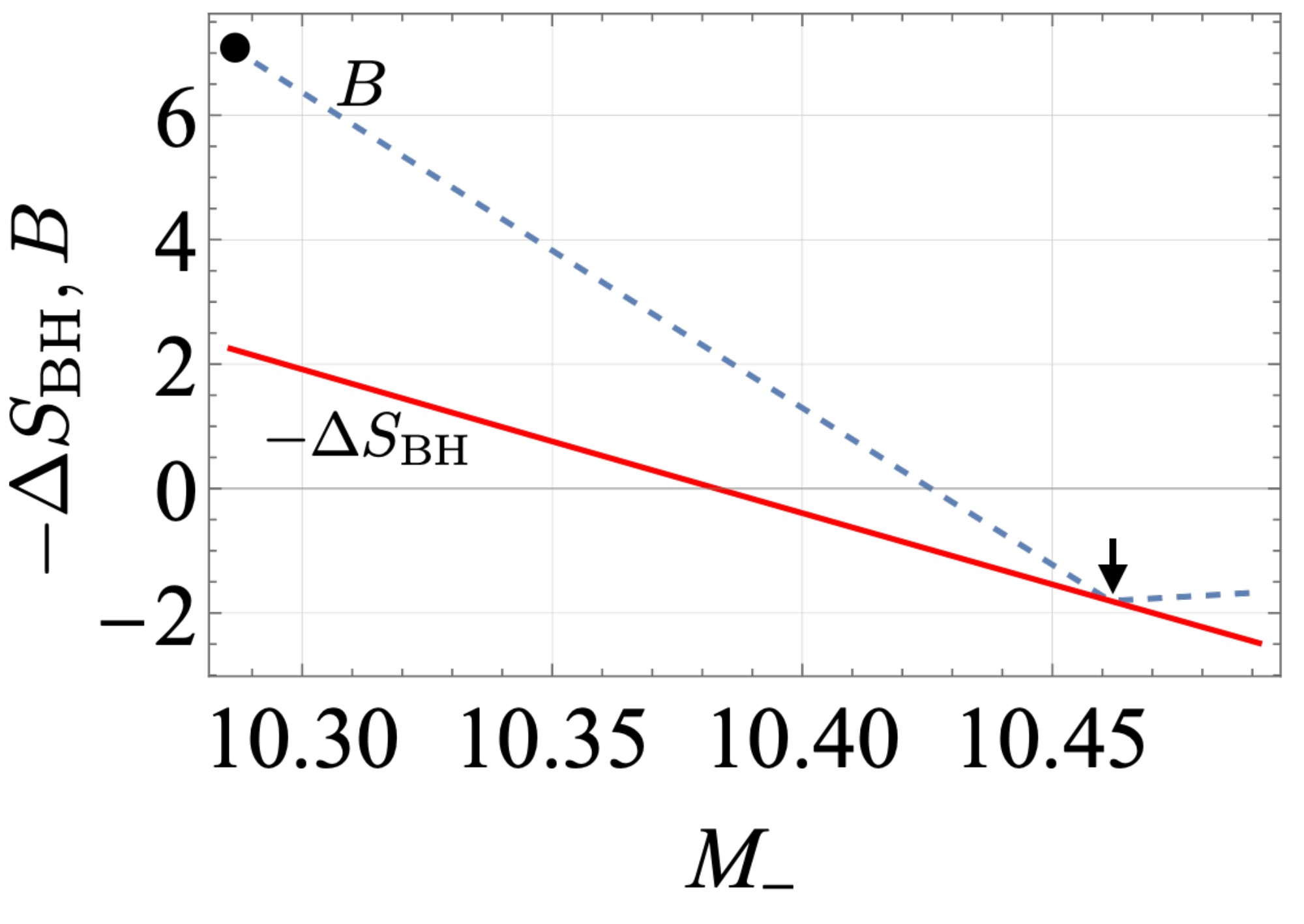}
  \caption{The action $B$ and $-\Delta S_{\rm BH}$ for $M_+ = 8$ and $a_+=1.5$. The black marker and arrow indicate the action for the stationary solution and the action minimum, respectively.
  }
\label{overall_plot}
   \end{figure}

In most cases, $\Delta S_{\rm BH}$ is negative and the Bekenstein-Hawking entropy decreases in the vacuum decay process. However, we find that for a rapidly spinning seed black hole, the $\Delta S_{\rm BH}$ can be positive and the semi-classical approximation is apparently broken (Figure \ref{overall_plot}). In such a case, we have $e^{-B} \gg 1$ and the standard prescription to estimate the transition amplitude in the Euclidean path integral is not valid due to the breakdown of the semi-classical approximation. However, in the context of the thermodynamics, this is nothing but the second law of thermodynamics, i.e., a preferred direction of a transition is determined so that entropy increases. Therefore, if the transition with $\Delta S_{\rm BH} > 0$ describes a mere thermal phase transition, it would be regarded as a physical process.

\section{Superradiant instability of a Kerr-AdS${}_5$ black hole}\label{QNM_part}
It is known that a rotating black hole in AdS${}_5$ is unstable at least against scalar perturbations, which is known as the superradiant instability. In this section, we briefly review the scalar quasi-normal (QN) modes for the \kdsf spacetime. We then compare the classical instability of QN modes with the quantum-mechanical instability of the metastable \kdsf background that we study in the present work. In the former part of this section, we introduce the QN modes of a \kdsf black hole. In the latter part, we then numerically compute the time scale of the superradiant instability and compare it with the lifetime of the false vacuum we estimated in the previous section.
\subsection{Scalar perturbations in \kdsf}
The Kerr-AdS${}_5$ black hole with different two spins has the metric
\begin{align}
\begin{split}
ds^2 &= - \frac{\Delta_r}{\rho^2} \left( dt - \frac{a_1 \sin^2 \theta}{1-a_1^2} d \phi - \frac{a_2 \cos^2\theta}{1-a_2^2} d \psi \right)^2 + \frac{\Delta_{\theta} \sin^2 \theta}{\rho^2} \left( a_1 dt - \frac{r^2+a_1^2}{1-a_1^2} d\phi \right)^2\\
&+ \frac{1+r^2}{r^2 \rho^2} \left( a_1 a_2 dt - \frac{a_2 (r^2 +a_1^2) \sin^2 \theta}{1-a_1^2} d\phi - \frac{a_1 (r^2 + a_2^2) \cos^2 \theta}{1-a_2^2} d\psi \right)^2 \\
&+ \frac{\Delta_{\theta} \cos^2\theta}{\rho^2} \left( a_2 dt - \frac{r^2+a_2^2}{1-a_2^2} d\psi \right)^2 + \frac{\rho^2}{\Delta_r} dr^2 + \frac{\rho^2}{\Delta_{\theta}} d\theta^2,
\end{split}
\end{align}
where the AdS radius is set to unity and
\begin{align}
\Delta_r &\equiv \frac{1}{r^2} (r^2+a_1^2) (r^2+a_2^2) (1+r^2) -2M=\frac{1}{r^2}(r^2-r_0^2)(r^2-r^2_-)(r^2-r^2_+),\\
\Delta_{\theta} &\equiv 1-a_1^2 \cos^2 \theta -a_2^2 \sin^2 \theta,\\
\rho^2 &\equiv r^2 + a_1^2 \cos^2 \theta+a_2^2 \sin^2 \theta.
\end{align}
The inner and outer horizon radii are denoted as $r_-$ and $r_+$, respectively, and $r_0$ is the imaginary root of $\Delta_r$.
Then we evaluate the superradiant instability with a scalar field $\Psi (t,r,\theta, \phi, \psi)$ with mass $\mu$. Let us start with the Klein Gordon equation,
\begin{equation}
\left[\nabla_{\mu} \nabla^{\mu} -\mu^2 \right] \Psi = 0,
\end{equation}
and decomposing $\Psi$ as $\Psi = e^{-i \omega t +im_1 \phi + i m_2 \psi} \Theta (\theta) \Pi (r)$, one has the following radial and angular equations:
\begin{align}
\begin{split}
& \frac{1}{r} \frac{d}{dr} \left( r \Delta_r \frac{d \Pi (r)}{dr} \right)  - \left[ \lambda + \mu^2 r^2 + \frac{1}{r^2} \left( a_1 a_2 \omega -a_2 (1-a_1^2)m_1 - a_1 (1-a_2^2) m_2 \right)^2  \right] \Pi (r)\\
&+ \frac{(r^2 + a_1^2)^2 (r^2 + a_2^2)^2}{r^4 \Delta_r} \left( \omega - \frac{m_1 a_1 (1-a_1^2)}{r^2+a_1^2} - \frac{m_2 a_2 (1-a_2^2)}{r^2+a_2^2}  \right)^2 \Pi (r) = 0,
\end{split}\label{radial_equation}\\
\begin{split}
&\frac{1}{\sin \theta \cos \theta} \frac{d}{d\theta} \left( \sin \theta \cos \theta \Delta_{\theta} \frac{d \Theta (\theta)}{d\theta} \right) - \left[-\lambda \omega^2 + \frac{(1- a_1^2) m_1^2}{\sin^2 \theta} + \frac{(1- a_2^2) m_2^2}{\cos^2 \theta} \right.\\
&\left. - \frac{(1-a_1^2) (1-a_2^2)}{\Delta_{\theta}} (\omega+m_1 a_1+m_2 a_2)^2 + \mu^2 (a_1^2 \cos^2\theta +a_2^2 \sin^2 \theta) \right] \Theta (\theta) =0,
\end{split}
\end{align}
where $\lambda$ is the separation constant that is determined so that $\Theta (\theta)$ is non-singular at the poles. For the equal spin case $a_1=a_2=a$, the angular equation reduces to hypergeometric differential equation, and the $\lambda$ is obtained in the analytic form
\begin{equation}
\lambda = (1-a^2) \left[ l (l+2) -2 \omega a (m_1+m_2) -a^2 (m_1+m_2)^2 \right] + a^2 \omega^2 +a^2 \Delta (\Delta-4),
\end{equation}
where $l=1,2...$ is the angular mode. Then we can get QN modes by solving radial equation satisfying the proper boundary condition, i.e., ingoing propagation at the outer horizon and $\Pi(r) \to 0$ at infinity. Performing the following transformations:
\begin{align}
r &\to z \equiv \frac{r^2- r_-^2}{r^2 - r_0^2}, z_0 \equiv \frac{r_+^2 -r_-^2}{r_+^2-r_0^2}\\
\Pi(r) &\to X(z) \equiv z^{\theta_-/2} (z-z_0) ^{\theta_+/2} (z-1)^{-\Delta/2} \Pi(z),
\end{align}
we see the angular equation (\ref{radial_equation}) reduces to the Heun's differential equation
\begin{align}
\frac{d^2 X}{dz^2} + \left[ \frac{1- \theta_-}{z} + \frac{-1 + \Delta}{z-1} + \frac{1- \theta_+}{z-z_0}  \right] \frac{d X}{dz} + \left( \frac{\kappa_1 \kappa_2}{z (z-1)} - \frac{K}{z (z-1) (z-z_0)} \right) X = 0,
\end{align}
with
\begin{align}
\theta_i &\equiv \frac{i\left(\omega - (m_1+m_2) \Omega_{i}\right)}{2\pi T_i},\
T_i \equiv \frac{r_i^2 \Delta_r'(r_i)}{4 \pi (r_i^2+a^2)^2}, \ 
\Omega_{i} \equiv \frac{a \Xi}{r_i^2 +a^2}, \\
\kappa_1 &\equiv - \frac{1}{2} \left( \theta_- + \theta_+ -\Delta - \theta_0 \right), \ 
\kappa_2 \equiv - \frac{1}{2} \left( \theta_- + \theta_+ -\Delta + \theta_0 \right),\\
\begin{split}
K &\equiv -\frac{1}{4} \left\{ \frac{\lambda + \mu^2 r_-^2 - \omega^2}{r_+^2 - r_0^2} + (z_0-1) [(\theta_+ + \theta_- -1)^2 -\theta_0^2-1] \right.\\
&~~~~~~~~~~~~~~~~~~~~~~~~~~~~~~~~~~~~~~~~~~~~~~~~~\left. +z_0 \left[ 2 (\theta_+-1) (1- \Delta) + (2-\Delta)^2 -2 \right] \right. \biggr\},
\end{split}
\end{align}
where the subscript $``i"$ is $i = -$, $+$, or $0$.
The general solution of the Heun's equation is defined around the singularity $z\sim 0, 1, z_0, \infty$, and we impose the ingoing boundary condition at the black hole horizon $r_+$ ($z=z_0$) and the Dirichlet condition at the AdS boundary $r=\infty$ ($z=1$). These boundary condition is expressed as
\begin{align}
\Pi (z) \sim
\begin{cases}
(z-z_0)^{-\theta_+/2} &\text{for} \ z\to z_0 \ (r \to r_+),\\
(z-1)^{\Delta/2} &\text{for} \ z\to 1 \ (r \to \infty).
\end{cases}
\end{align}
Then the local solutions at $z=z_0$ and $z=1$ that satisfy the boundary conditions are
\begin{align}
X \sim
\begin{cases}
 H \ell \left( \frac{z_0}{z_0-1}, \frac{-K}{z_0-1}; \kappa_1, \kappa_2, 1-\theta_+, \Delta-1; \frac{z_0-z}{z_0-1} \right)&\text{for} \ z\sim z_0 \ (r \sim r_+),\\
 H\ell \left( 1-z_0, \kappa_1 \kappa_2 -\tilde{K}; \kappa_1, \kappa_2, \Delta-1, 1-\theta_-; 1-z \right)&\text{for} \ z\sim 1 \ (r \sim \infty),
\end{cases}
\end{align}
where $H\ell$ is the Heun function. In this paper we used $\it{Wolfram\  Mathematica}$ to search for QN frequencies $\omega_{lm_1m_2 n}$. The resolution test was performed in our former work \cite{Koga:2022vun}. 

\subsection{Superradiant instability vs. vacuum metastability in \kdsf}
\begin{figure}[htbt]
\centering
\includegraphics[width=12cm]{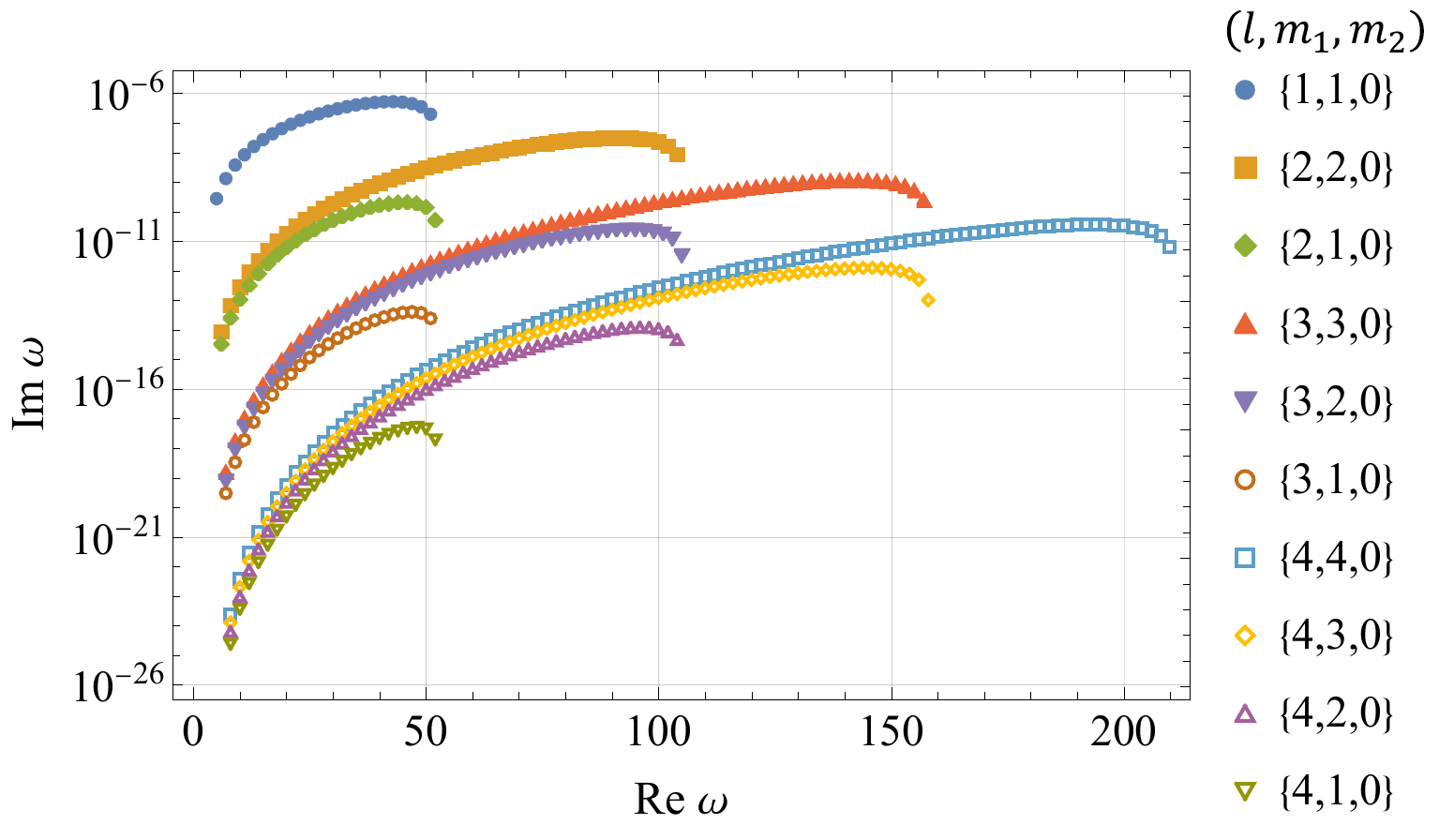}
\caption{Plot of unstable QN modes for the angular modes of $l=1$, $2$, $3$, and $4$. One can see that $l=1$ mode has the largest value of $\text{Im}(\omega_{lm_1m_2n})$. We fix the background metric with $a=0.001$ and $M=10^{-5}$, and the mass of the scalar field is set to $\mu=0.01$. Note that the AdS radius is set to unity in this plot.}
\label{QNMplot}
\end{figure}

\begin{figure}[htbt]
\centering
\includegraphics[width=9cm]{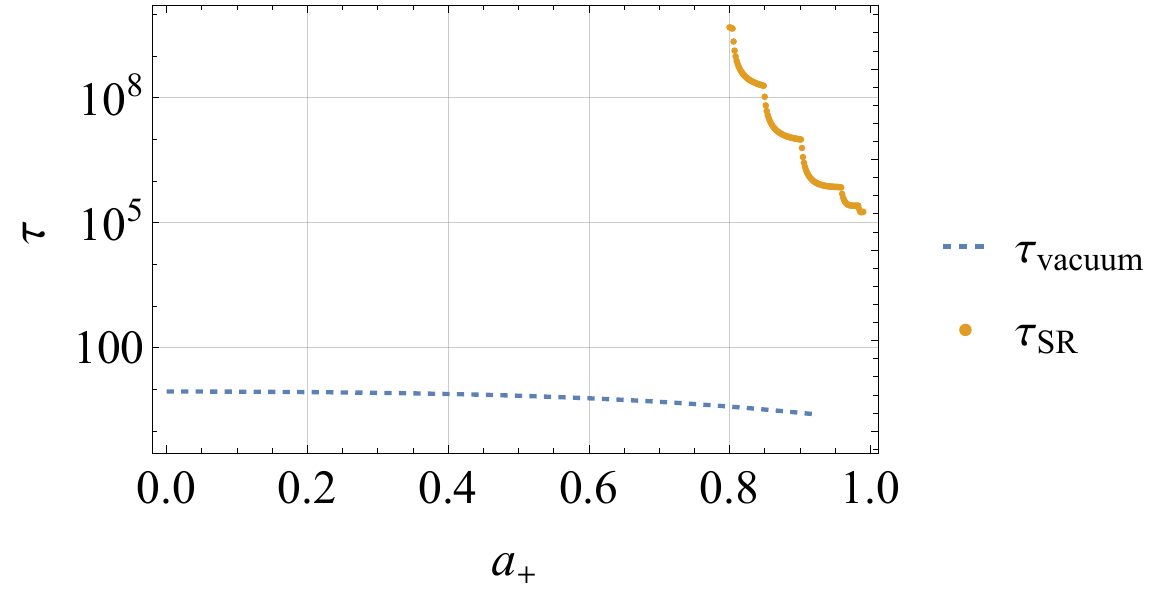}
\caption{Plot of $\tau_{\rm SR}$ and $\tau_{\rm vacuum}$ with respect to the spin parameter $a_+$. We set $\mu=0.01$, $M_+=2$ and $a_+ \leq 0.92$. The value of $\tau_{\rm vacuum}$ is computed for the most probable process of vacuum decay whose decay amplitude is determined only by $\Delta S_{\rm BH}$.
}
\label{lifetime}
\end{figure}

The superradiant instability of the \kdsf black hole appears when it has a QN frequency $\omega = \omega_{lm_1m_2n}$ whose imaginary part is positive as the amplitude of each QN mode is $\sim e^{\text{Im} (\omega_{lm_1m_2n}) t}$. If $\text{Im}(\omega_{lm_1m_1n})>0$, the background spacetime is unstable against the perturbations. The time scale of the superradiant instability, $\tau_{\rm SR}$, is estimated as
\begin{align}
\tau_{\rm SR} = \frac{1}{\max (\text {Im}(\omega_{lm_1m_2n}))},
\end{align}
where $\max (\text {Im}(\omega_{lm_1m_2n}))$ is the maximum positive value of $\text {Im}(\omega_{lm_1m_2n})$ among all QN modes.
It is known that for small black holes, $r_+ \ll 1$, with rapid rotations, there exist unstable QN modes whose frequencies satisfy the following inequality
\begin{align}
\text {Re}(\omega_{lm_1m_2n})<(m_1+m_2)\Omega_{+},
\end{align}
where $\Omega_+ \equiv a(1-a^2)/(r_+^2 + a^2)$. This condition means that if the unstable modes exist, higher angular modes have a wider frequency band of unstable QN modes. To see this, we plot QN modes for $l=1,2,3,4$ in Figure \ref{QNMplot}.
Then we compare the superradiant instability with the vacuum metastability in \kdsf spacetime. The lifetime of the false vacuum state in the \kdsf background, $\tau_{\rm vacuum}$, is
\begin{align}
\tau_{\rm vacuum}=\frac{1}{\Gamma}, \quad\Gamma=Ae^{-B},
\end{align}
where $A$ is the pre-factor that originates from the zero modes of the instanton and loop corrections to the saddle point solutions, and $B$ is the Euclidean action of a stationary solution. We usually determine the pre-factor with the dimensional analysis as the magnitude of transition amplitude is mainly governed by the exponential factor. We here assume $A=1/r_+$ as the size of a seed black hole determines the typical scale of a nucleated bubble \cite{Gregory:2013hja}. 
We set the \kdsf background with $M_+=2, l_+=7, l_-=4$ and set the mass of the scalar field to $\mu=0.01$. We then compare the two lifetimes in Figure \ref{lifetime}. It shows that the superradiant instability is absent in the low-spin region ($a \lesssim 0.8$), but it appears for rapid spins ($a \gtrsim 0.8$). The superradiant instability is significant, i.e., $\tau_{\rm SR}$ is shorter, for rapid rotations. On the other hand, the more rapid the rotation of the seed \kdsf black hole is, the longer $\tau_{\rm vacuum}$ is. From Figure \ref{lifetime}, one can see that at least in the parameter set, the superradiant instability does not interrupt the nucleation process of a vacuum bubble\footnote{Although other instabilities originating from gravitons and photons may contribute to the superradiant instability, the qualitative feature of superradiance is similar for other species of particles.}. Note that we do not exclude the possibility that the superradiant instability dominates the false vacuum state with $\tau_{\rm SR} \lesssim \tau_{\rm vacuum}$ in different parameter sets.

\section{Discussion and Conclusion}
\label{conclusions}
In this paper, we studied the nucleation process of a vacuum bubble in the \kdsf spacetime. It is an extension of the cosmological scenario \cite{Banerjee:2018qey,Banerjee:2019fzz,Banerjee:2020wix}, in which a quasi-dS${}_4$ expansion is realized on the nucleated bubble in an AdS${}_5$ spacetime, to a situation where a gravitational impurity, i.e., a \kdsf black hole, is involved in the false vacuum state. It is known that an AdS spacetime is non-linearly unstable and small black holes may form in the AdS due to the gravitational turbulent instability \cite{Bizon:2011gg,Dias:2011ss}. We then found that the cosmological scenario works even in the less symmetric background, i.e., \kdsf false vacuum state, without contradicting the de Sitter swampland conjectures.

We here computed the transition amplitude, $e^{-B}$, of the bubble nucleation process with the Euclidean path integral technique. We then found that there exists a parameter region that admits the bubble nucleation in the \kdsf spacetime, and that the most probable process is given by a stationary solution for which the bubble wall has no oscillation in the imaginary time and the transition amplitude is governed by the change of the Bekenstein-Hawking entropy $e^{-B} = e^{\Delta S_{\rm BH}}$. We also found that for a rapidly rotating and massive seed black hole, the Bekenstein-Hawking entropy in the system increases due to the bubble nucleation and leads to $e^{-B} =e^{\Delta S_{\rm BH}} \gg 1$. It can be naively regarded as the breakdown of the semi-classical approximation. On the other hand, one could interpret it as a thermal transition with the increment of the entropy based on the generalized second law of thermodynamics \cite{Bekenstein:1972tm,Bekenstein:1973ur,Bekenstein:1974ax}. In that sense, we may admit the parameter region leading to $e^{\Delta S_{\rm BH}} \gg 1$ as a parameter region where a vacuum bubble is nucleated by the thermal activation of the seed black hole (see e.g., Ref. \cite{Oshita:2016btk}).

As the seed black hole has the rotation while it is confined in the AdS barrier, the false vacuum we considered is not even classically stable due to the superradiant instability. Then we compare the lifetime of the false vacuum state determined by the vacuum bubble nucleation $\tau_{\rm vacuum}$ and that associated with the superradiance $\tau_{\rm SR}$. We found that there exists a parameter set for which $\tau_{\rm SR} \gg \tau_{\rm vacuum}$. Although we do not exclude the possibility that the superradiant instability disturbs the nucleation of vacuum bubbles in the \kdsf spacetime, we show that the cosmological scenario can be realized even with the superradiance.


\section*{Acknowledgement}
The authors thank Shinji Mukohyama, Toshifumi Noumi, Kota Ogasawara, and Hiroaki Tahara for their helpful comments. This work was supported in part by Japan Society for the Promotion of Science (JSPS) Grant-in-Aid for Research Activity Start-up Grant No. 21K20371 (NO) and JSPS Fellows Grant No. 20J22946 (KU). NO is also supported by the Special Postdoctoral Researcher (SPDR) Program at RIKEN and the FY2021 Incentive Research Project at RIKEN.

\appendix
\section{Derivation of ${\cal L}_m^{\rm (on-shell)} = -(\sigma -\Delta P)$ in (\ref{wall_action})}
\label{app_onshell}
The off-shell action density of the fluid on the bubble $s_w$ is
\begin{equation}
s_w = \sqrt{-q} {\cal L}_m = \sqrt{-q} \left( -\sigma - \varphi u_{(i}\xi_{j)} q^{ij} + (1/2) C_{ij} q^{ij} \Delta P \right),
\end{equation}
where $q = {\rm det} (q_{ij})$ and
a projection $C_{ij}$ can be expressed by $C_{ij}dx^i dx^j \equiv -d\tau^2 + h^2(R)(d\psi + A_a dx^a)^2$, which is obtained by tracing out $\hat{g}_{ab}$ from the induced metric $q_{ij}$ in (\ref{induced_metric_qij})\footnote{As shown in Ref. \cite{Gibbons:2004uw}, the odd-dimensional Kerr spacetime, say $(2n+1)$-dimensions, with equal rotations has a $(2n-1)-$sphere as a part of the metric. It can be rewritten as the Hopf fibration over a base space $\mathbb{CP}^{n-1}$. In our case ($n=2$), the traced part of $\hat{g}_{ab}$ is associated with the base space of $S^2$ and may be irrelevant to the black hole rotation. On the other hand, the remnant spacial part $(d\psi + A_a dx^a)^2$ corresponds to the azimuthal angle or the rotations of the black hole, i.e., $(d\psi + A_a dx^a) = \sum_{i} \mu_i^2 d\phi_i$ (for more details and the notations of $\mu_i$ and $\phi_i$, see Appendix B in Ref. \cite{Gibbons:2004uw}). Remember that we redefined $\psi$ to obtain the corotating frame in (\ref{psi_redefine}). As such, the projection $C_{ij}$ can be interpreted as a projection onto the rotation (and proper-time) directions. We then conclude that the last term in the Lagrangian contributes to the anisotropic pressure of the fluid caused by the rotation. Indeed, $\Delta P$ is zero when there is no rotations.}. Note that we here assume $P = -\sigma$.
The energy momentum tensor is
\begin{align}
{\cal S}_{ij} = \frac{-2}{\sqrt{-q}} \frac{\delta s_w}{\delta q^{ij}} = -2 \frac{\partial {\cal L}_m}{\partial q^{ij}} + q_{ij} {\cal L}_m,
\end{align}
and one can easily check this reduces to the expression of ${\cal S}_{ij}$ in (\ref{emt_wall}).
The on-shell Lagrangian ${\cal L}_m^{\rm (on-shell)}$ is
\begin{equation}
{\cal L}_m^{\rm (on-shell)} = \left( -\sigma - \varphi u_{(i}\xi_{j)} q^{ij} + C_{ij} q^{ij} R^2 \Delta P \right)_{\rm on-shell} = - \sigma + \Delta P,
\end{equation}
where we used
\begin{align}
(u_{(i}\xi_{j)} q^{ij})_{\rm on-shell} &= 0, \label{varphi_emt}\\
(C_{ij} q^{ij})_{\rm on-shell} &= \frac{1}{R^2}.
\end{align}
(\ref{varphi_emt}) holds and $\varphi$ vanishes in ${\cal L}_m^{\rm (on-shell)}$ as we use the co-rotating coordinates. Still, the rotation effect remains in the geometrical quantity (e.g., centrifugal force) in the total on-shell action.

\end{document}